\documentclass{JHEP3}
\usepackage{graphicx}
\usepackage{epsfig,amsmath,amsthm}
\newcommand{\be}{\begin{equation}}
\newcommand{\ee}{\end{equation}}
\newcommand{\bea}{\begin{eqnarray}}
\newcommand{\eea}{\end{eqnarray}}
\def\bse{\begin{subequations}}
\def\ese{\end{subequations}}

\def\IZ{\relax\ifmmode\hbox{Z\kern-.4em Z}\else{Z\kern-.4em Z}\fi}

\newcommand{\non}{\nonumber \\}

\def\half{\frac{1}{2}} \def\quart{\frac{1}{4}}

\def\del{{\partial}}

\def\td{\tilde{d}}

\def\vnabla{\vec{\nabla}}
\def\vA{\vec{A}}
\def\tgamma{\widetilde{\gamma}}
\def\htt{{\hat t}}
\def\hi{{\hat i}} \def\hj{{\hat j}} \def\hk{{\hat k}}

\def\presub{\vspace{.5cm} \noindent}

\def\bi{\begin{itemize}} \def\ei{\end{itemize}}

\def\Schw{Schwarzschild }
\def\({\left(} \def\){\right)}
\def\[{\left[} \def\]{\right]}

\title{ \center{Einstein's action and the harmonic gauge in terms of Newtonian fields}}

\author{Barak Kol\\
Racah Institute of Physics, Hebrew University\\
Jerusalem 91904, Israel\\
{\tt\href{mailto:barak_kol@phys.huji.ac.il}{barak\_kol@phys.huji.ac.il}}
}
 \author{
Michael Smolkin\\
Perimeter Institute for Theoretical Physics\\
 Waterloo, Ontario N2L 2Y5, Canada \\
 \email{msmolkin@perimeterinstitute.ca}
 }

\abstract{The ``Newtonian'' or non-relativistic decomposition of
Einstein's gravitational field is useful in the post-Newtonian
approximation. We obtain the full non-quadratic Einstein-Hilbert
action in terms of these fields as well as the harmonic gauge fixing term 
and find fairly simple expressions. We discuss alternatives to the harmonic gauge.}

\begin{document}

\section{Introduction}

Newton's universal law of gravitation \cite{Principia} can be
expressed through the action of a (static) gravitational potential
$\phi$ \be
 S =  -\frac{1}{8 \pi G} \int dt\, d^3 x\, \(\vec{\nabla}
 \phi\)^2 \label{Nfield}  \ee
together with its coupling to massive objects $\int dt \(
\frac{m}{2}\, \dot{\vec{r}}^{~2}
 - m\, \phi(\vec{r}) \).$

In Einstein's theory of gravity (General Relativity, GR) \cite{GR}
the gravitational field is promoted to a space-time metric
$g_{\mu\nu}$ subject to the Einstein-Hilbert action  \be
 S_{EH}=-\frac{1}{16 \pi G} \int \sqrt{-g}\, d^4x\, R[g]  ~,
  \label{EH} \ee
 where the overall sign in (\ref{EH}) reflects the signature
convention for $g_{\mu\nu}$ which we take to be time-like. The
coupling to massive objects becomes $-m \int d\tau$ , where
$x^\mu(\tau)$ is the particle's trajectory, the proper time is
defined by $d\tau^2:=g_{\mu\nu}\, dx^\mu\, dx^\nu$, and for
clarity we used $c=1$ units.

In the Newtonian limit Einstein's gravity reduces to Newton's.
Yet, Einstein's theory calls for a 10 component field, while
Newton's theory has only one. A natural question arises:
\emph{what is the physical role of the other metric components in
the Newtonian limit?}

In 2007 we proposed a change of variables from the metric to
certain non-relativistic gravitational (NRG) fields based on a
temporal Kaluza-Klein reduction. One of them can be identified
with the Newtonian potential $\phi$, and the others also have  a
role, even if subleading, in this limit, as we review below, and
hence they can be called ``Newtonian''. However, the action (and
hence also the equations of motion) were not fully known.
\emph{Determining the full action including gauge fixing terms is the objective of this paper.}

We carried the computation using a non-orthonormal frame within
Cartan's method, namely a hybrid method which incorporates both a
non-trivial frame and a non-trivial metric. The process turned out
to be related to existing work in the Kaluza-Klein literature
\cite{AulakhSahdev,Maheshwari} but it was not used so far in the
gravitational post-Newtonian context to the best of our knowledge.
We also computed the harmonic gauge fixing term for these fields,
thereby completing the evaluation of the total action.


\presub {\bf Background}. The Newtonian limit of Einstein's
gravity was an issue since its very beginning \cite{GR,LorentzDroste1917}, continued with \cite{EIH} and developed into \cite{FockInfeld}. The
interest further arose in the last three decades with the
development of the Post-Newtonian (PN) approximation, see
\cite{BlanchetRev,Schaefer-account} for recent reviews and
\cite{Blanchet:1984uw,BlanchetDamour89} for some earlier work. The central
application of the Post-Newtonian approximation is the analysis of
the two body problem in GR and the resulting gravitational wave
signature. This problem is crucial for the global effort to
directly detect gravitational waves \cite{IFOs} and eventually
interpret their signals.

In 2004 the effective field theory (EFT) approach to GR
\cite{GoldbergerRothstein1} was put forward. This approach borrows
from the ideas of effective quantum field theories. It is based on
a hierarchy of scales in the problem and essentially replaces the
traditional method of dealing with finite size objects through
matched asymptotic expansion by introducing instead effective
interactions of point particles with their background. See \cite{DamourFarese} for early precursors of the EFT method.

In \cite{CLEFT-caged,NRG} we introduced the NRG fields \footnote{See \cite{BlanchetDamour89,Schaefer-gravitomagnetic} for early precursors of the NRG fields.}
 and used
them to give what is probably the shortest derivation of the
leading post-Newtonian correction, known as the
Einstein-Infeld-Hoffmann interaction \cite{EIH}. Their utility
received strong support from \cite{GilmoreRoss} who reproduced the
2PN effective action (see \cite{BlanchetRev} and references
therein) through the EFT approach. The authors of
\cite{GilmoreRoss} compared the NRG fields against the standard
metric, and found the NRG fields to be better suited for the
calculational purpose. After the first arXiv version of this paper appeared further
support was received when the computation was successfully
extended to reproduce order 3PN using an impressive computerized
calculation \cite{FoffaSturani}. The computations in
\cite{NRG,GilmoreRoss,FoffaSturani} required the evaluation of
certain terms in the NRG action and those were evaluated in a
perturbative weak field expansion and sometimes with the help of
computerized computation. The full and concise non-perturbative expressions presented here
allows to readily expand and read off the gravitational vertices
for the PN perturbation theory and \emph{it should be useful for
further research including advancing the post-Newtonian
state-of-the-art and reaching order 4PN} (which is not available
yet in any method), a goal announced in \cite{FoffaSturani}. In
such higher orders the recursive diagrammatic relations of
\cite{dressed} should be helpful as well. Related interesting and
relatively recent work on PN and/ or EFT appeared in
\cite{related}.

This paper is organized as follows. In section
\ref{fields-section} we review the definition of NRG fields. In
section \ref{action-section} we present the Einstein-Hilbert
action in terms of these fields (the derivation is relegated to an
appendix). In section \ref{gauge-section} we discuss possible
gauge choices from the PN perspective and compute the harmonic
gauge fixing term in NRG fields.

\section{NRG fields}
\label{fields-section}

The non-relativistic limit is defined as the limit of slow
velocities (relative to the speed of light). As a zeroth
approximation we consider the stronger time-independent
(stationary) limit. In this limit it is natural to perform a
temporal Kaluza-Klein (KK) dimensional reduction
\cite{CLEFT-caged,NRG} (the original and standard KK reduction
\cite{Kaluza-Klein} was spatial, of course)
 \be
 ds^2 = e^{2 \phi}(dt - A_i\, dx^i)^2 -e^{-2 \phi}\, \gamma_{ij}\,
 dx^i dx^j ~. \label{KKansatz} \ee
This relation defines a change of variables from $g_{\mu\nu}$ to
$(\phi,A_i,\gamma_{ij}), ~i,j=1,2,3$ which we call
``non-relativistic gravitational fields'', or in short
``NRG-fields''.

The action, when translated into NRG-fields and within the
stationary limit becomes \be
 S = -\frac{1}{16\pi G} \int dt\, d^{3}x \sqrt{\gamma}
  \[ -R[\gamma] + 2\, \left| \vnabla \phi \right|^2 -  \frac{1}{4}\, e^{4\phi} F^2 \]~,\label{stat-action} \ee
 where $\left| \vnabla \phi \right|^2 = \gamma^{ij}\, \del_i \phi\,
\del_j \phi$ and  $F^2$ is conventionally defined by $F^2=F_{ij}
F^{ij}, ~~F_{ij}=\del_i A_j- \del_j A_i$.

The above-defined NRG fields (\ref{KKansatz}) resemble the
definition of the well-known ADM fields \cite{ADM} (see also the
review \cite{3+1rev} and references therein) given by $ds^2= N^2
dt^2-\gamma_{ADM~ij} (dx^i+N^i dt) (dx^j+N^j dt)$, where $N,N^i$
are the lapse and shift. Moreover, for $A_i=0=N^i$ the definitions
coincide as they do at linear order for all fields (after a
suitable Weyl-rescaling of the ADM fields).  However, there is a
marked difference between the two: the ADM fields are designed for
the initial value problem and the time evolution of the metric,
while NRG fields are designed for the post-Newtonian evolution of
say a two-body problem. The NRG fields are defined by a temporal
KK reduction, while ADM is nothing but a spatial KK reduction (in
order to focus on the time evolution), namely in NRG  the shift is
applied to time and in ADM to space. More concretely this means
that NRG fields transform nicely under time-independent coordinate
reparameterizations, while ADM transforms nicely in the
space-independent case. So in this sense ADM and NRG are sort of
opposites actually.

After the first arXiv version of this paper appeared a detailed
and practical comparison between NRG, ADM and a modified ADM
fields was made \cite{NRG-ADM} in the PN context and NRG was shown
to be the more economical field definition starting at order 2PN.


It is interesting that even though the Kaluza-Klein (KK) ansatz
was available since 1921, it was not applied in full to the
post-Newtonian approximation before \cite{CLEFT-caged,NRG} as far as
we are aware. One possible reason is that the usual KK procedure
is for a compact space-like dimension while here it is applied to
time.

\presub {\bf Physical meaning of NRG fields $\phi, A_i,~\gamma_{ij}$}. 
Comparing the kinetic term for $\phi$ in the
action (\ref{stat-action}), with the Newtonian field action
(\ref{Nfield}) together with the respective interaction terms we
find it natural to identify $\phi$ with the Newtonian potential.
To leading order around flat-space time (Newtonian limit) this definition coincides with the accepted definitions of the Newtonian potential $\phi_N$ in the literature, for instance $g_{00} \simeq 1 + 2 \phi_N$.
Due to the nearly stationary nature of the Newtonian limit, we find $\phi$ to be a good way to extend this definition beyond the linear order.

The vector potential $\vA$ has an action which apart for an
overall sign resembles the magnetic part of the 4d Maxwell action,
and accordingly it is natural to call $F$ the gravito-magnetic
field and call $\vA$ the gravito-magnetic vector potential. This
name originates in a certain similarity between gravity and
electro-magnetism. The strong similarity between Newton's
gravitational force and Coulomb's static electrical force,
together with the observation that the transition from
electro-statics to electro-dynamics requires to supplement the
scalar electric potential by a vector potential, promoted already
in the 19th century suggestions to add a vector potential to the
gravitational degrees of freedom.  In fact, it is known how to
obtain such a vector potential in the weak gravity/ Post-Newtonian
approximation to GR, a point of view known as
``Gravito-Electro-Magnetism (GEM)'' (see for example
\cite{wikiGEM,Schaefer-gravitomagnetic}) and references within).
Just like in electromagnetism the vector potential couples at
leading order to the charged current, which in gravity is the mass
current, namely $S \supset m \int dt \vec{v} \cdot \vec{A}$, and it
 is responsible for a gravito-magnetic force between two mass currents. 
A spinning object consists of a dipole of mass current and
accordingly it couples universally to the magnetic field $\vec{B}=\vec{\nabla} \times \vec{A}$ 
through $S \supset \half \int dt \vec{J} \cdot \vec{B}$. 
We note the reversed sign of the kinetic term for $F$.
This is directly related to the fact that in gravity both the
current-current and the spin-spin force have an opposite sign
relative to electro-dynamics, and so ``north poles attract''
\cite{Wald-spin}.

Finally, starting with order 2PN we must also account for the
3-metric tensor $\gamma_{ij}$ which comes with a standard
Einstein-Hilbert action in 3d (this is achieved through the Weyl
rescaling factor in front of $\gamma_{ij}$ in the ansatz). This
field has no electromagnetic analogue.

\section{NRG action}
\label{action-section}

In this section we go beyond the stationary approximation of the last section and
derive the full, time-dependent action for the
NRG fields. For greater generality we work in an arbitrary
space-time dimension $d$. We perform the change of variables in
two steps. In the first we replace \be
 g_{\mu\nu} \to (\phi,\vA,\tgamma_{ij}) \ee and in the second step \be 
 \tgamma_{ij} \to \gamma_{ij} ~. \nonumber \ee 

The first step is achieved through a dimensional reduction \be
 ds^2 = e^{2\phi} \( dt - A_i\, dx^i \)^2 - \tgamma_{ij} dx^i dx^j \label{dim-red-ansatz}
\ee
 The Einstein-Hilbert action $S=-1/(16 \pi G) \int R dV$
becomes \bea
 S &=& -\frac{1}{16 \pi G} \int e^\phi\, \sqrt{\tgamma}\, d^{d-1}x\, dt
 \non
 && \left\{ -
 \quart e^{2 \phi} \bar{F}^2
 -\quart \, e^{-2\phi} \( \dot{\tgamma}_{ij} \dot{\tgamma}_{kl} \tgamma^{ik}
 \tgamma^{jl} - \(\tgamma^{ij} \dot{\tgamma}_{ij}\)^2 \) - \bar{R}[\tgamma]
 \right\} \label{dim-red-action}
 \eea
where we define \bea
 D_i &:=& \del_i + A_i \del_t \non
 \bar{F}_{ij} &:=& D_i A_j - D_j A_i = F_{ij} + A_i \dot{A}_j - A_j
 \dot{A}_i ~, \label{def-Di} \eea
 a dot denotes a time derivative and $\bar{R}[\tgamma]$ denotes the Ricci scalar of the spatial metric
 $\tgamma_{ij}$ where the derivatives in its expression are replaced everywhere as follows $\del_i \to
 D_i$. Borrowing notation from the {\it Mathematica} software this definition
 can be stated by \be
 \bar{R}[\tgamma] := R[\tgamma] ~ /. ~ \del_i \to D_i ~.\ee

Note the following alternative forms for two of the terms in
(\ref{dim-red-action}): $\dot{\tgamma}_{ij} \dot{\tgamma}_{kl}
\tgamma^{ik} \tgamma^{jl} = -\dot{\tgamma}^{ij}
\dot{\tgamma}_{ij}$ and $\tgamma^{ij} \dot{\tgamma}_{ij} = \del_t
\log \det \tgamma$.

The action (\ref{dim-red-action}) for the dimensionally reduced
fields (\ref{dim-red-ansatz}) is the essential step towards
incorporating time dependence into the action, and we find that it
leaves quite a compact result: the $\bar{F}^2, ~\bar{R}$ terms
generalize appropriately the well-known stationary action, while
the remaining term is a kinetic term for $\gamma_{ij}$ which
defines an associated metric on field space, namely on the space
of metrics, and it coincides with one of the deWitt metrics
\cite{deWitt-metric}.
In order to compute the action we used a non-orthonormal frame
within Cartan's method, namely a hybrid method which incorporates
both a non-trivial frame and a non-trivial metric (see appendix
\ref{derivation-section}).
 
The second step is a Weyl rescaling \be
 \tgamma_{ij} = e^{-2\phi/\td} \gamma_{ij} \label{Weyl-rescale}
 \ee
 where we use a shorthand notation \be
 \td:= d-3 ~. \label{def-td} \ee
Altogether the metric reads now \be
 ds^2 = e^{2\phi} \( dt - A_i\, dx^i \)^2 - e^{-2\phi/\td} \gamma_{ij} dx^i dx^j \label{NRG-ansatz} \ee
 which agrees with (\ref{KKansatz}) for $d=4$. The Weyl rescaling is designed
such that the stationary action for $\gamma_{ij}$ will be the
standard Einstein-Hilbert action, without the $e^\phi$ prefactor
present in (\ref{dim-red-action}). We denote the perturbation of
the spatial metric $\gamma_{ij}$ by \be
 \sigma_{ij} := \gamma_{ij}- \delta_{ij} ~. \label{def-sigma} \ee

Substituting into the action we obtain our main result, \emph{the fully non-linear and
time-dependent Einstein-Hilbert action for the NRG fields} \bea
 S &=& -\frac{1}{16 \pi G} \int \sqrt{\gamma}\, d^{d-1}x\, dt
 \non
 && \left\{ \(1+\frac{1}{\td}\) \left| D_i  \phi \right|^2 -
 \quart e^{2(1+1/\td)\phi} \bar{F}^2 - \bar{R}[\gamma]  \right. \non &&
 +2(1+1/\td) \gamma^{ij}\, \dot{A}_i\,  D_j \phi \non &&
 -\quart \, e^{-2(1+1/\td)\phi} \( \dot{\gamma}_{ij} \dot{\gamma}_{kl} \gamma^{ik}
 \gamma^{jl} - \(\gamma^{ij} \dot{\gamma}_{ij}\)^2 \)  \non  & & \left.
 -e^{-2(1+1/\td)\phi} \( (1+\frac{1}{\td})\, \dot{\phi}\, \gamma^{ij}\dot{\gamma}_{ij} - \frac{(\td+2)(\td+1)}{\td^2} \dot{\phi}^2 \)  \right\}
 \label{post-Weyl-action} \eea

In particular for $d=4$ space-time dimensions the full NRG or
Post-Newtonian action is \bea
 S &=& -\frac{1}{16 \pi G} \int \sqrt{\gamma}\, d^3x\, dt
 \non
 && \left\{ 2 \left| D_i  \phi \right|^2 -
 \quart e^{4\phi} \bar{F}^2 - \bar{R}[\gamma]  \right.
 + 4 \gamma^{ij}\, \dot{A}_i\,  D_j \phi \non && \left.
 -\quart \, e^{-4 \phi} \( \dot{\gamma}_{ij} \dot{\gamma}_{kl} \gamma^{ik}
 \gamma^{jl} - \(\gamma^{ij} \dot{\gamma}_{ij}\)^2 \)
 -e^{-4\phi} \( 2 \, \dot{\phi}\, \gamma^{ij}\dot{\gamma}_{ij} - 6 \dot{\phi}^2 \)  \right\}
 \label{4d-PN-action} ~. \eea

We tested our expression for the action in several limits. In the
stationary limit it reduces correctly to (\ref{stat-action}). A
strong test is provided by the sector with two time derivatives.
Since it contains no spatial derivatives, the same sector can be
computed by a performing dimensional reduction over space,
defining fields similar to the ADM decomposition. The action is
given by an the analogue of (\ref{stat-action}) which is then
followed by a change of variables to NRG. The result matches with
the expansion of the two time derivatives of
(\ref{dim-red-action}) given by (\ref{eqn:S2}).

The expression for the action, together with the gauge fixing term
 (\ref{Hgauge-fix})
 conveniently encodes the Feynman rules for the diagrammatic computation of the two-body effective
action in the EFT method
\cite{GoldbergerRothstein1,NRG,GilmoreRoss,dressed}. The relevant
terms up to order 2PN include the quadratic time-dependent
vertices \be \dot{\phi}^2, ~ \dot{A}^2 \ee and the non-quadratic
vertices of the form \be \phi\, \dot{\phi}^2,~ \nabla \phi\,
\dot{\phi}\, A, ~ \phi\, (\nabla A)^2, ~ \sigma_{ij}\, \del_i
\phi\, \del_j \phi ~.\ee
 Our expression, together with the recursive relations of \cite{dressed} should be useful for computing higher PN orders.


\subsection{Massive Kaluza-Klein action}

Suppose we take the KK direction to be space-like (as in the
original KK idea) and denote it by $z$. Then our action
(\ref{post-Weyl-action}) is related to the full action including
$z$-dependent fields, namely the massive modes. It turns out that
this action is essentially known in the literature: the 4+1
pre-Weyl rescaling action in compact form (analogous to our
(\ref{dim-red-action}) ) was found by Aulakh and Sahdev
\cite{AulakhSahdev} and an expanded 4+1 post Weyl rescaling
expression (analogous to our (\ref{eqn:EH_KKdecomp}) ) was found
by Maheshwari \cite{Maheshwari} (see also the reviews
\cite{KKrev}). Below we illustrate how to use our result to obtain
the full $4+1$ action after Weyl rescaling and in compact form.

The metric ansatz becomes \be ds^2 = e^{-2\phi/\td}
\gamma_{\mu\nu} dx^\mu dx^\nu - e^{2\phi} \( dz - A_\mu\, dx^\mu
\)^2 \ee where $\mu,\nu=0,1,\dots,d-2$. We observe that the
substitution \be
 t \leftrightarrow z \ee
amounts to reversing the Ricci scalar (since it reverses the
metric in the non-orthonormal frame) \be
 R[g_{\mu\nu}] \to -R[g_{\mu\nu}] ~.\ee
This allows us to obtain the action by reversing
(\ref{post-Weyl-action}). For concreteness let us present the form
of the KK action for $d=5$ \bea S &=& \frac{1}{16 \pi G} \int
\sqrt{-\gamma}\, d^3x\, dt\, dz
 \non
 && \left\{ \frac{3}{2} \left| D_\mu  \phi \right|^2 -
 \quart e^{3\phi} \bar{F}^2 - \bar{R}[\gamma]  \right.
 + 3 \gamma^{\mu\nu}\, D_\mu \phi\, \del_z A_\nu\,   \\ && \left.
 -\quart \, e^{-3 \phi} \( \gamma^{\mu\rho}
 \gamma^{\nu\sigma} (\del_z \gamma_{\mu\nu}) (\del_z \gamma_{\rho\sigma})  - \(\gamma^{\mu\nu} \del_z \gamma_{\mu\nu}\)^2 \)
 -e^{-3\phi} \( \frac{3}{2} \, (\gamma^{\mu\nu}\del_z \gamma_{\mu\nu}) \del_z{\phi}\,  - 3 (\del_z \phi)^2 \)
 \right\} \nonumber
 \label{4d-KK-action} . \eea
This expression could be useful for studying non-linear
interactions involving massive KK modes.

\section{The gauge}
\label{gauge-section}

In computing the two-body effective action we are free to choose
the gauge (choice of coordinates). It affects both the length of
the calculation as well as the end result since the effective
action depends on the coordinate location of the two bodies and as
such is not gauge invariant.

The harmonic gauge is used throughout the PN literature. It is
defined by  \be
 0=\Gamma^\mu := \Gamma^\mu_{\nu\rho}\, g^{\nu\rho} \equiv -\frac{1}{\sqrt{-g}}\, \del_\nu \( \sqrt{-g} g^{\mu\nu} \) \ee
where $\Gamma^\mu_{\nu\rho}$ is the Christoffel symbol for the
space-time metric $g$. The corresponding gauge fixing action is taken to be \be
 S_{GF} = \frac{1}{32 \pi G} \int d^dx \sqrt{g}\, g_{\mu\nu} \Gamma^\mu\,
 \Gamma^\nu  \label{S-GF-gen} \ee

In the first subsection we evaluate (\ref{S-GF-gen}) in NRG fields
(\ref{NRG-ansatz}). In the second subsection we critically discuss the rationale for
choosing the harmonic gauge. We note that the harmonic gauge is
special for being covariant under Lorentz transformations, while
the PN approximation breaks the symmetry between time and space
and therefore choosing the harmonic gauge is not obvious. We
proceed to analyze this question explicitly order by order in the
PN expansion up to 2PN. We summarize our results and discuss them
at the end of that subsection.

\subsection{Harmonic gauge fixing term}

In order to express the harmonic gauge fixing term (\ref{S-GF-gen}) in terms of NRG fields we find in appendix \ref{derivation-section} how to
express $\Gamma_\mu$ in terms of a frame and a frame metric
(\ref{harmonic-hybrid}). Substituting
(\ref{frame},\ref{frame-metric}) we find
 \bea
 S_{GF} &=& \frac{1}{2 \cdot 16 \pi G} \int \sqrt{\gamma}\, d^{d-1}x\, dt \label{Hgauge-fix}  \\
 & & \left\{ \( e^{(1+1/\td) \phi}\, \gamma^{ij} D_i A_j + 2 \(1+\frac{1}{\td}\)\, e^{-(1+1/\td) \phi}\, \dot{\phi} - \half e^{-(1+1/\td) \phi}\, \del_t \log \gamma \)^2 -
  \left| \bar{\Gamma}_i[\gamma] - \dot{A}_i \right|^2 \right\}   \nonumber \eea
 where we define \bea
 \bar{\Gamma}_{ijk}[\gamma] &:=& \Gamma_{ijk}[\gamma] ~ /. ~ \del_i \to D_i \non
     &=& \Gamma_{ijk}[\gamma] +\half \(A_j \dot{\gamma}_{ik} + A_k \dot{\gamma}_{ij} - A_i \dot{\gamma}_{jk} \) ~. \label{def-Gamma-bar}  \eea


The expression (\ref{Hgauge-fix}) was tested by using two
additional frame representations of the metric: an
\emph{orthonormal} frame and another which shifts the field $\phi$  from the
frame to the metric.

\subsection{Comparing gauges for the PN limit}

In general, the freedom to choose a gauge could be a blessing
provided that it is chosen judiciously. Actually given a specific
problem there must be an \emph{optimal gauge} (simply because the
computational cost should be bounded from below and non-zero).
Here we are considering the PN approximation, which defines a
certain class of problems, meaning that only partial information
about the system is known. Accordingly we expect that there would
not be a single optimal gauge, but rather an optimal class of
gauges where some of the gauge freedom is fixed while some
residual gauge freedom remains to fit a given, specific problem.

In this subsection we ask \emph{what is the optimal gauge (class)
for the post-Newtonian approximation}.

The harmonic gauge fixing appears to be standard in the
literature.\footnote{Sometimes one encounters the closely related
linearized harmonic gauge.} However, the rationale for that is not
clear to us. One of the appealing features of the harmonic gauge
(perhaps characterizing it) is its covariance property being a
vector under global Lorentz transformations of (asymptotically)
flat space-times. Yet, in the PN approximation, Lorentz invariance
is broken to spatial rotations by the choice of a rest frame
(choice of time), and hence we proceed to consider a more general
class of gauges with less symmetry order by order in the PN
expansion.

{\bf The Newtonian order}. The newtonian field is gauge-invariant
to leading, stationary gauge transformations. Accordingly the question of gauge choice is
moot.

{\bf Quadratic stationary gauge fixing starting from 1PN}. In the
PN approximation we decompose the gravitational field into the
Newtonian potential $\phi$, the gravito-magnetic vector $A_i$ and
perturbations $\sigma_{ij}$ of the spatial metric tensor
$\gamma_{ij}$ (\ref{NRG-ansatz},\ref{def-sigma}). The vector
appears in the stationary, unperturbed action (\ref{stat-action})
in a Maxwell form, and accordingly the standard Lorentz gauge
fixing term must be chosen at the quadratic level \bea S_{GF} &=&
\frac{1}{32\pi G} \int d^dx\, C_0^{~2} \non
 C_0 &=& \vec{\nabla} \cdot \vec{A}  ~. \label{C01} \eea
 in order to avoid the propagator from mixing the different vector components.
 This gauge fixing term affects order 1PN through the
diagram of gravito-magnetic interaction (vector exchange).

Similarly $\gamma_{ij}$, the spatial metric has an
Einstein-Hilbert action and for simplicity of the propagator we
better choose the harmonic gauge (note: only in the spatial
directions), at least linearly. This gauge choice shows up first
at order 2PN through the diagram of $\sigma$ exchange.

{\bf Quadratic but non-stationary sector starting from 1PN.} Before gauge fixing the
action includes a term of the form $\dot{\vec{A}} \cdot
\vec{\nabla} \phi$ just like the post-Coulomb approximation to
electro-magnetism. This term induces a mixing 2-vertex for $\phi,
\vec{A}$.
 A generic gauge fixing term includes also a $\dot{\phi}^2$ vertex
and hence generically three diagrams contribute: an exchange of
$\phi$ with a retardation vertex, a $\phi$ -- $A$ exchange diagram
and a diagram with a $\phi$ -- $A$ -- $\phi$ chain.

However, after integration by parts the $\phi-A$ 2-vertex is
proportional to $\dot{\phi} \vec{\nabla} \vec{A}$ and can be
removed by shifting the gauge condition on $\vec{\nabla} \vec{A}$,
more specifically by deforming the gauge condition (\ref{C01}) to
become \be
 C_0 = \vec{\nabla} \cdot \vec{A} + 2 \(1+\frac{1}{\td}\) \dot{\phi}  ~. \label{C02}\ee
 This constraint turns out to be part of the harmonic gauge 
and hence \emph{up to order 1PN the harmonic gauge is certainly
optimal}. We note that just like the electro-magnetic two-body
effective action is gauge invariant at any post-Coulomb order, the
same holds in the first order of the gravitational case, due to
the similarity of the actions up to this order, and hence the sum
of the three diagrams is independent of gauge.

Similarly, at order 2PN with harmonic gauge the diagram of $A$
exchange with a retardation vertex replaces two diagrams with a
$\sigma \dot{A}$ mixing which would appear in a generic gauge.

{\bf Non-Quadratic gauge fixing starting from order 2PN}. We still
have freedom to deform the gauge constraint (\ref{C02}) with
non-linear terms. This freedom can be used to cancel any term in
the action which is proportional to the linearized gauge
constraint.

\begin{figure}[t!]
\centering \noindent
\includegraphics[width=8cm]{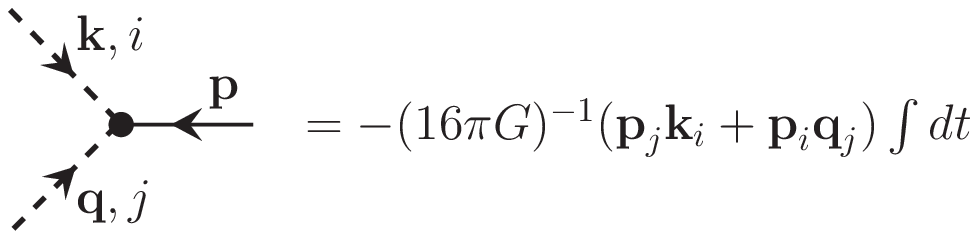}
\caption[]{Diagrammatic representation of the new static
interaction obtained after gauging away $A^i\nabla_i\phi\dot\phi$
from the Einstein-Hilbert action. } \label{fig:new-static-vertex}
\end{figure}

At order 2PN the vertex $\vec{\nabla}\phi\, \dot{\phi} \vec{A}$ 
 appears in two diagrams: 5(m) and 5(n) of \cite{GilmoreRoss}.
  We may cancel this term by deforming the gauge condition as
follows
 \be
C_0 =  \vec{\nabla} \cdot \vec{A} + 2 \(1+\frac{1}{\td}\)
\dot{\phi} + A^i\del_i\phi ~. \label{C03}\ee
 At the same time this gauge fixing produces a term of the form
$\vec{\nabla}\phi\,  \vec{A}\, \vec{\nabla}\vec{A}$ which at first
sight is already present (see eq. (35) in \cite{GilmoreRoss}),
however the tensor structure of the new term is different and one
must add the
 static $Y$-type interaction presented in figure
\ref{fig:new-static-vertex}.

\begin{figure}[t!]
\centering \noindent
\includegraphics[width=8cm]{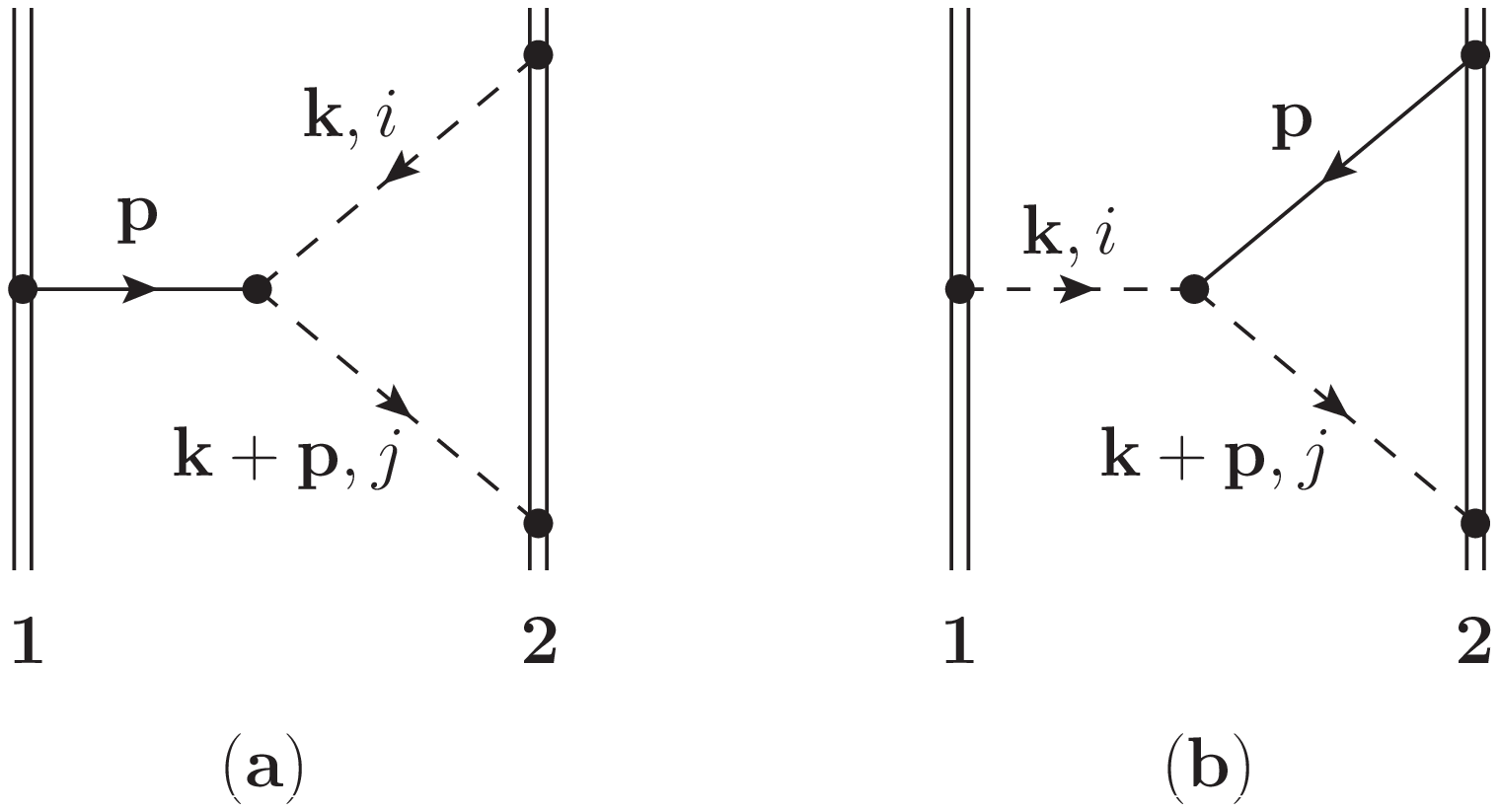}
\caption[]{New static Feynman diagrams obtained after gauging away
$A^i\nabla_i\phi\dot\phi$ from the Einstein-Hilbert action. (a)
and (b) replace figures 5(l) and 5(k) of Gilmore-Ross
respectively.} \label{fig:gauge-change1}
\end{figure}

Computing the value of the terms added to the diagrams we find
 \bea
 \mathrm{Fig.}\ref{fig:gauge-change1}(a)&=&{d-3 \over d-2}(16\pi G m_2)^2\,{m_1 \over 2}\int dt \int {d^{\,d-1}k \over (2\pi)^{d-1}}\,{d^{\,d-1}p \over (2\pi)^{d-1}}
 \,{(v_2\cdot p)(v_2\cdot k) \over p^2k^2(p+k)^2}\,e^{i p \cdot r}
 ~,
 \non
 \mathrm{Fig.}\ref{fig:gauge-change1}(b)&=&-{d-3 \over d-2}(16\pi G m_2)^2\,{m_1 \over 2}\int dt \int {d^{\,d-1}k \over (2\pi)^{d-1}}\,{d^{\,d-1}p \over (2\pi)^{d-1}}
 \,{(v_1\cdot p)(v_2\cdot p) \over p^2k^2(p+k)^2}\,e^{i k \cdot r}
 ~.
 \eea
Integrating over one of the wave numbers and evaluating the
resulting Fourier transform by means of (A.2) and (A.10) of
\cite{dressed} respectively, yields (in d=4)
 \bea
 \mathrm{Fig.}\ref{fig:gauge-change1}(a)&=&-2{G^2m_2^2m_1 \over r^2}\Big[\vec v_2^2-2(\vec v_2\cdot\hat r)^2\Big]
 ~,
 \non
 \mathrm{Fig.}\ref{fig:gauge-change1}(b)&=&{G^2m_2^2m_1 \over r^2}\Big[ 3 (\vec v_1\cdot\hat r) (\vec v_2\cdot\hat r)-(\vec v_1\cdot\vec v_2)\Big]
 ~.
 \eea

We conclude that while in new gauge (\ref{C03}) avoids computing
two diagrams (fig. 5(m,n) of \cite{GilmoreRoss}) it generates two
new terms in the diagrams in fig. 5(k,l) which appear to be of
comparable computational difficulty. Hence it is not clear that
the new gauge (\ref{C03}) is any better than the harmonic.

\presub {\bf Comments about 3PN}. At order 3PN log terms appear,
reflecting an RG flow
\cite{BlanchetRev,GoldbergerRothstein1,FoffaSturani}. The
associated divergences are canceled by available redundant terms
on the world-line which reflect the choice of coordinates, but not
in observables. Presumably this is a price for working with
harmonic coordinates. Indeed, the equation for changing the radial
coordinate of the \Schw solution from standard to harmonic has log
solutions even in asymptotically flat space, and those terms could
be essential in background which are not asymptotically flat.

\presub {\bf Optimal Post-Newtonian gauge: summary and
discussion.} We critically analyzed the optimal choice of gauge
for the PN approximation. We found that the harmonic gauge is
certainly optimal at the level of propagators and 2-vertices
(starting with 1PN). At 2PN we found a different gauge which
reduces the number of diagrams. However, this does not lead to a
reduction in computational difficulty. The reason for that is that
each standard Feynman diagram may represent several computational
terms corresponding to the positioning of derivatives and to
various index structures. A faithful diagrammatic representation
of the computational difficulty requires to distinguish vertices
of a field from those of its derivatives and also to reflect the
index structure. Hence altogether we do not find a gauge which is
better than harmonic, at least up to 2PN.

Our analysis suggests that \emph{the harmonic gauge could be the
optimal PN gauge} (in the bulk). In hindsight, the reason for that
could be simple. The total PN action is composed of the bulk
gravitational action and the interaction with the object's
world-line. The computational difficulty is dominated by bulk
irreducible loops, which depend on bulk vertices while non-linear
world-line vertices contribute only factorizable diagrams (see for
example \cite{dressed}). Hence the choice of gauge depends only on
the bulk action which enjoys an increased symmetry (relative to
PN). The harmonic gauge can be characterized by its background
independence, namely in any background the quadratic action for
the fluctuations is given simply by the Lichnerowicz operator.
Probably this characterization leads to its advantage also for
non-quadratic orders around flat space.


\subsection*{Acknowledgements}

It is a pleasure to thank M. Duff, R. Emparan, G.
Gibbons and G. Sch\"{a}fer for correspondence regarding the
manuscript. Special thanks to K. Hinterbichler who attracted our attention to \cite{Maheshwari}.

This research is supported by The Israel Science Foundation grant
no 607/05, by the German Israel Cooperation Project grant DIP
H.52, and by the Einstein Center at the Hebrew University. Research at
Perimeter Institute is supported by the Government of Canada through Industry Canada and
by the Province of Ontario through the Ministry of Research \& Innovation.

\appendix
\section{Deriving the NRG action}
\label{derivation-section}

In this section we shall derive (\ref{dim-red-action}) -- the full Einstein-Hilbert action in terms of the
dimensionally reduced fields (\ref{dim-red-ansatz}). We use a non-orthonormal frame within
Cartan's method, namely a hybrid method which incorporates both a
non-trivial frame and a non-trivial metric, see for example \cite{MTW}.

The dimensionally reduced fields (\ref{dim-red-ansatz}) define a natural \emph{frame of 1-forms} \bea
 \theta^{\htt} &=& e^\phi \( dt - A_i\, dx^i \)  \non
 \theta^{\hi} &=& dx^i \label{frame} \eea
and \emph{a metric} \be
 g_{ab}=\mbox{diag}(1,-\tgamma_{ij}) \label{frame-metric} \ee
 where $a,b=(\htt,\hi)$ are frame indices and
from hereon we shall replace $\tgamma \to \gamma$ throughout the derivation for clarity of notation and without risking confusion.

It is convenient to record the inversion of (\ref{frame}) \bea
 dt   &=& e^{-\phi} \theta^{\htt} + A_i \theta^{\hi} \non
 dx^i &=& \theta^{\hi}\label{inverted-frame} \eea
 as well as the dual basis of vectors \be
 \left\{ \begin{array}{rcl}
  e_{\htt} &=& e^{-\phi} \del_t \non
  e_{\hi} &=& \del_i + A_i \del_t \equiv D_i \\ \end{array} \right.
    \qquad
  \left\{ \begin{array}{rcl}
    \del_t &=& e^\phi\, e_{\htt} \non
    \del_i &=& e_{\hi} - e^\phi A_i e_{\htt} \\ \end{array} \right.  \ee

\emph{The connection 1-forms} $\omega^a_{~b}$ are the solution of \bea
 d\theta^a + \omega^a_{~b} \theta^b &=& 0 \non
 \omega_{ab} + \omega_{ba}          &=& dg_{ab} ~. \eea

In order to solve this system one defines the components of $\omega^a_{~b}$ to be \be
  \omega^a_{~b} = \omega^a_{~bc} \theta^c \ee
  and these components are given by the sum \be
  \omega_{abc} = \Sigma_{abc} + \Gamma_{abc} \label{omega-Sigma-Gamma} \ee
  where $\Sigma_{abc}$ is related to the torsion of the frame,  $\Gamma_{abc}$ is the Christoffel symbol for the frame metric $g_{ab}$
  (it is symmetric in $bc$, namely $\Gamma_{a[bc]}=0$) and both are defined and computed below.

First one defines a tensor $C_{ab}^{~~c}$ by \be
 d\theta^c + \half C_{ab}^{~~c}\, \theta^a \theta^b =0 ~. \ee
 $C_{ab}^{~~c}$ is antisymmetric in $ab$, namely $C_{(ab)}^{~~~c}=0$.
In our case one has \bea
 C_{\hi \hj}^{~~\htt} &=& e^\phi \bar{F}_{ij} \non
 C_{\htt \hi}^{~~\htt} &=& D_i \phi + \dot{A}_i \eea
 where $\bar{F}, D_i$ are defined in (\ref{def-Di}).

$\Sigma_{abc}$ is defined by \be
 \Sigma_{abc} = \half \( C_{abc} - C_{bca} - C_{cab} \) ~. \ee
$\Sigma_{abc}$ is also antisymmetric in $ab$, namely
$\Sigma_{(ab)c}=0$, and in our case we have \bea
 \Sigma_{\hi \hi \htt} &=& \Sigma_{\hi \htt \hj} = \half e^\phi \bar{F}_{ij} \non
 \Sigma_{\htt \hi \htt} &=& D_i \phi + \dot{A}_i \label{Sigma} \eea

Next one computes the Christoffel symbol defined as usual by \be
 \Gamma_{abc}= \half \( \del_b g_{ac} + \del_c g_{ab} - \del_a g_{bc}  \) ~.\ee
In our case it is given by \bea
 \Gamma^{\hi}_{~\hj \hk} &=& \bar{\Gamma}[\gamma]^{i}_{~jk}[\gamma] \non
 \Gamma^{\htt}_{~\hi \hj} &=& \half e^{-\phi} \dot{\gamma}_{ij} \non
 \Gamma^{\hi}_{~\htt \hj} &=& \half e^{-\phi}\, \gamma^{ik} \dot{\gamma}_{kj} ~, \label{Gamma} \eea
 where $\bar{\Gamma}$ is defined by (\ref{def-Gamma-bar}).
Substituting back (\ref{Sigma},\ref{Gamma}) into (\ref{omega-Sigma-Gamma}) provides us with the connection $\omega_{abc}$.

\emph{The Einstein-Hilbert action} is expressed in terms of the
curvature, which is related to derivatives of the connection.
However, after integration by parts we found that it could be
expressed in terms of the connection only, so there is no need to
compute the curvature \be
 (-16 \pi G)\, S_{EH} = \int R dV \simeq \int \sqrt{g}\, \theta\,  \( \omega^{abc} \omega_{abc} - \omega_a \hat{\omega}^s \) \label{omega-action} \ee
 where $\simeq$ denotes equality up to boundary terms and we define \bea
 \theta &:=& \prod_a \theta^a \non
 \omega_a &:=& g^{bc}\, \omega_{abc} = g^{bc} \Gamma_{abc} - C_{ab}^{~~a} \non
 \hat{\omega}_a &:=& \omega^{a}_{~ba} = \Gamma^{a}_{~ba} + C_{ab}^{~~a} \eea
 We note that if one were to consider $\int f dV R$ where $f$ is any scalar function,
the integration by parts would add the term
$-(\omega^a-\hat{\omega}^a) \del_a \log f$.

Substituting the values of $\omega_{abc}$ into
(\ref{omega-action}), we encounter the expression \be -\(
\bar{\Gamma}^{ijk} \bar{\Gamma}_{kij}-\bar{\Gamma}^j_{~ij}\,
\gamma^{kl} \bar{\Gamma}^{i}_{~kl} + (\gamma^{ij}
\bar{\Gamma}^k_{~jk} - \gamma^{kl} \bar{\Gamma}^i_{kl}) ((D_i \phi
+ \dot{A}_i)\) \ee
 where everywhere the connection here is of the spatial metric alone $\Gamma_{ijk}=\Gamma[\gamma]_{ijk}$. Recalling that the measure is $e^\phi\, \sqrt{\tgamma}\, d^{d-1}x\, dt$ and using the comment at the end of the last paragraph this expression can be integrated back by parts to yield $-\bar{R}[\gamma]$.
Finally replacing back from the temporary notation of this section $\gamma \to \tgamma$ we obtain the expression for the  action (\ref{dim-red-action}) that we sought to derive.

\presub {\bf Harmonic gauge fixing term}. In order to calculate
this term it is useful to express the harmonic constraint in terms
of a hybrid frame formalism. We obtain the following relation \be
 \Gamma_a[g_{\nu\rho}] :=  e_a^\mu\, \Gamma_\mu[g_{\nu\rho}] =
 \Gamma_a[g_{bc}] + \Gamma_a^\theta
\label{harmonic-hybrid} \ee
 where the first term is the usual
expression only in terms of the metric in the frame basis \be
 \Gamma_a[g_{bc}] := g^{bc} \( \del_b g_{ac} -\half \del_a g_{bc}
 \) \ee
 while the second term involves also the frame \be
  \Gamma_a^\theta := \( g_{ab}\, g^{\nu\rho} +
  e^\rho_a\, e^\nu_b  - e^\nu_a\, e^\rho_b \)  \del_\nu
  \theta^b_\rho \ee

\section{NRG action expanded}

In this section we present expressions for the expanded action and
group it according to the number of time derivatives for the
purposes of the PN approximation. These expressions were computed
using a rather different method (relying on a presentation of the
metric by an \emph{orthonormal} frame) and were confirmed to be
identical with the expanded form of (\ref{post-Weyl-action}).

Since each term in the scalar curvature contains precisely two
derivatives of the metric with respect to either a spatial or a
temporal coordinate, one can organize all the terms into three groups according to the
number of temporal derivatives
 \be
 S_{EH}(g) = \frac{1}{16\pi G}\int R[g] \rightarrow S^{(0)}(\gamma,A,\phi)+S^{(1)}(\gamma,A,\phi)+S^{(2)}(\gamma,A,\phi)
 ~.
 \label{eqn:EH_KKdecomp}
 \ee

We find
 \bea
 S^{(0)}(\gamma,A,\phi) &=& - \frac{1}{16\pi G} \int d^{\,d}x
 \sqrt{\gamma}\[ -R[\gamma] + \frac{d-2}{d-3} \(\del \phi\)^2  -
 \frac{1}{4}e^{2(d-2)\phi/(d-3)}F^2 \] \label{eqn:S0-S1}
 \\
 S^{(1)}(\gamma,A,\phi)&=&\frac{1}{16\pi G}\int d^{\,d}x \sqrt{\gamma}
 \Big[-{1\over 2}\,e^{2(d-2)\phi/(d-3)}F^{ij}\dot A_{[i}A_{j]}+{\del_t(\gamma\,\gamma^{ij})\over\gamma}\,\nabla_i A_j
 \nonumber
 \\
 &&\quad\quad\quad\quad\quad\quad\quad\quad\quad\quad\quad\quad\quad\quad\quad~
 -2 \, {d-2 \over d-3}\nabla^i\phi \left( \dot A_i + \dot\phi\,A_i \right) \Big]
 ~, \nonumber
 \eea
where
a dot denotes the derivative with respect to time, the metric
$\gamma$ is being used to raise and lower the indices such as
$\(\del \phi\)^2 = \gamma^{ij}\, \del_i \phi\, \del_j \phi$ and we
use the standard definitions $F^2=F_{ij}$ and
$F^{ij},~~F_{ij}=\del_i A_j- \del_j A_i$. Finally,
$S^{(2)}(\gamma,A,\phi)$ is given by

 \bea
 S^{(2)}(\gamma,A,\phi) &=& \frac{1}{16\pi G} \int d^{\,d}x
 \sqrt{\gamma} ~ {1 \over 4}
 \left( A^2 -  e^{-2(d-2)\phi/(d-3)} \right)
 \Big( (\gamma^{ij}\dot\gamma_{ij})^2
 +\dot\gamma^{ij}\dot\gamma_{ij} \Big)
 \non
 &-&\frac{1}{16\pi G} \int d^{\,d}x \sqrt{\gamma} ~ {1 \over 2} \left[
 \dot\gamma^{ij}\dot\gamma_{ik} A_j A^k
 + \gamma^{kl}\dot\gamma_{kl} \dot\gamma_{ij} A^i A^j \right]
 \non
 &+&\frac{1}{16\pi G} \int d^{\,d}x \sqrt{\gamma} ~
 {1 \over 2} \left[ \gamma^{ij}\dot A_i \dot A_j A^2 - \left(A^i \dot A_i \right)^2 \right]
 e^{2(d-2)\phi / (d-3)}
 \non
 &-&\frac{1}{16\pi G}\,{d-2 \over d-3}  \int d^{\,d}x \sqrt{\gamma} ~
 \, \dot\phi^2 \left[ A^2 + {d-1 \over d-3} \, e^{- 2(d-2)\phi / (d-3)} \right]
 \non
 &+&\frac{1}{16\pi G}\,{d-2 \over d-3} \int d^{\,d}x \sqrt{\gamma}\,e^{-2(d-2)\phi/(d-3)}
 \,\dot\phi \, \gamma^{ij}\dot\gamma_{ij}
 \non
 &+&\frac{1}{16\pi G} \int d^{\,d}x \sqrt{\gamma}
 \left[ \gamma^{kl}\dot\gamma_{kl} A^j \dot A_j
 +\dot\gamma^{ij}\dot A_i A_j \right]
 \non
 &-&\frac{1}{8\pi G} ~ {d-2 \over d-3} \int d^{\,d}x \sqrt{\gamma} ~
 \dot\phi \, A^i \dot A_i  \, ,
 \label{eqn:S2}
 \eea
with $A^2=\gamma^{ij} A_i A_j$ .


\end{document}